\documentstyle[epsfig]{article}
\begin{document}

\title{Phase offsets between core and conal components of
     radio pulsars and their emission altitudes}
\author{R.C.Kapoor\thanks{email : rck@iiap.ernet.in}\\
 Indian Institute Of Astrophysics, \\Koramangala,
 Bangalore, 560 034, India  \and
       C. S. Shukre\thanks{email : shukre@rri.res.in}\\
 Raman Research Institute, \\ Sadashivanagar,
 Bangalore, 560 080, India}

\date{\today}
\maketitle
\newpage
\begin{abstract}
 We present a new and a potentially powerful method for investigating
 emission altitudes of radio pulsar core and conal components by
 attributing them different altitudes. We provide a framework for a
 systematic understanding of resulting longitude offsets between them
 which are frequently observed. By investigating the contributions to
 these
 offsets due to aberration and the magnetic field line sweepback, we show
 that they are always dominated by aberration for all emission altitudes
 and inclination angles. This directly leads to the conclusion that the
 core emission does not necessarily come from the surface.
 Based on our results, the trends seen in the observational phase offsets
 imply that for a large number of pulsars the emission altitudes of core
 and conal components are close when compared to the light cylinder radius
 but not necessarily relative to the stellar radius. The altitude
 difference between core and conal components that we find are typically
 larger than the individual altitudes ascribed to them so far. Our
 results also allow very widely different but related core/cone altitudes.
 We find that data supports also this circumstance for some pulsars, which
 suggests a novel and natural explanation of the precursors in the Crab
 and similar pulsars. The pre- and post-cursor nature of these components
 arises because of large offsets caused by correspondingly large
 differences in the core and conal emission altitudes.
 We also show that the question of emission altitudes can not be divorced
 from considerations of the core/cone components' filling factors of the
 polar flux tube. We propose an empirical '1/3 rule' to concisely
 describe the observed core/cone morphologies. Combined with the
 core/cone phase offsets it allows a glimpse into the variation of
 these filling factors with altitude. Lastly, we outline how the full
 predictive potential of our results can be realised by combining them
 with a detailed analysis of the polarization and multifrequency
 observations.
\end{abstract}

\section{Introduction}
 Knowledge of the emission altitudes of radio pulses is of great
 importance for an understanding of the pulsar emission physics.
 Various methods for estimating emission altitudes have been used
 in the past (for a review see Cordes, 1992). These methods have been
 applied in Rankin (1990, 1993), Phillips (1992), and also by
 Nowakowski (1994),  Kijak and Gil (1998), Gangadhara and Gupta (2001),
 and Mitra and Rankin (2001), to derive altitudes for many pulsars. The
 latest method by Blaskiewicz et al. (1991) (henceforth BCW) proposed
 and utilised a novel offset between centroids of the pulse intensity
 profile and the linear polarization position angle curve, which would
 result due to aberration. Recently this method has been further used
 by von Hoensbroech and Xilouris (1997) (henceforth HX). However, a
 consensus on and an order in derived emission altitudes has not
 emerged. Here we propose another independent method to gain knowledge
 of altitudes of core and conal components, by attributing the
 observed phase offsets between them to different altitudes.

 The pioneering and sustained study of Rankin (Rankin 1983, 1990, 1993
 hereafter respectively R1, R4, and R6) has established that the mean
 pulse profiles of radio pulsars often consist of components designated
 as core and conal having differing characteristics. On the basis of a
 large body of data concerning the pulse widths, it has been suggested
 that the core emission emanates from the full polar cap at an altitude
 $r_{core}$ which is the same as the stellar radius, $R_*$ (usually
 taken as 10 km) (R4). The conal emission is inferred to come from
 a hollow cone at an altitude, $r_{cone}$, of about 10-20 $R_*$ (R6).
 This was the first instance when different emission altitudes were
 attributed to different pulse components. About the values of
 altitudes there is no unanimity at present (Kijak and Gil
 1998), but it is clear that core and conal emission altitudes could
 generally be different. Keeping in view the previously quoted
 emission altitudes of upto $200 R_*$, we take for them a general
 maximum figure of the order of $20 \%$ of the light cylinder radius
 $r_L = (c\,P/2\,\pi)$, although the altitudes could vary with the
 pulsar period $P$, the inclination angle between the magnetic and
 rotation axes $\alpha$, and the fraction of the emission cap that is
 active. However, later we also comment on the possibility of having
 arbitrarily large emission altitudes.

 It is obvious that components of pulsar radiation which emanate from
 different altitudes will show some differences in properties related
 to their emission altitudes. One such is the often seen non-coincidence
 of centres of the core components and mid-points of the conal pairs.
 This phase offset between core and conal components through kinematic
 effects provides a new method to learn about these emission
 altitudes. As we demonstrate later, it is unencumbered by the ignorance
 of details of the emission mechanism and allows interpretation of
 observations in more detail so as to constrain the theory.

 What follows is a framework for a systematic understanding of these
 phase offsets between core and conal components as being due to their
 differing altitudes. This approach has many advantages
 because one can learn about emission altitudes by focussing on
 {\it different} pulse components observed at {\it one} frequency
 rather than studying {\it one} component at {\it different}
 frequencies. Firstly this
 complements multifrequency considerations (Philips, 1992; Kijak and
 Gil, 1998). More importantly it allows us to skirt the complications
 arising from our ignorance of radius to frequency mapping and related
 time delays etc. (Cordes 1992).
 Furthermore, we expect that our study will clarify
 whether the phase offsets are as expected from only the contributions
 we are considering or call for additional contributions which provide
 us with hints about the details of the emission process. Lastly let us
 mention that core emission coming from the stellar surface has
 implications for the pulsar matter's equation of state leading to
 strong indications that pulsars are strange stars rather than neutron
 stars (Kapoor and Shukre 1999, 2001). This makes it very important to
 verify if core components truly emanate on the stellar surface. This
 interpretation of the remarkable pulse width relation of Rankin rests
 crucially on the assumption that core emission comes from the
 {\it full} polar cap (R4). We shall see what implications this has
 for the core/cone offsets.

 In the relation between the emission altitudes of core/cone components
 and the phase offsets due to these differing altitudes, one inevitable
 kinematical contribution is due to differential aberration (Kapoor and
 Shukre 1998 hereafter KS). Another contribution would be due
 to the magnetic field line sweepback (henceforth {\it mfs}), first
 considered by Shitov(1983, 1985). Although these two have been considered
 previously, most often separately, we here combine them properly for the
 first time, while keeping in mind the core/cone distinction. In addition
 to these two there would be dynamical contributions related to details
 of the emission process which are at present not known. Ignoring these
 latter contributions, we investigate what the geometrical considerations
 of pulsar beams as delineated by the open field line structure of an
 oblique dipole rotator can tell us about these longitude offsets and
 their relation to the locations of the core and conal emission regions
 in the context of the polar cap model of pulsar emission.

 For this purpose we adapt the general formulae of KS to the present
 situation. We then combine these with {\it mfs} results of Shitov.
 A net phase offset is then derived in the next section.
 Thus we treat aberration exactly and {\it mfs} in its presently
 available approximate form. Section 3 describes the behaviour of
 calculated offsets as the emission altitude is varied. Secton 4
 gives a comparison of the trends in the emission altitudes emerging
 from our results and observations, especially those which
 hitherto appeared puzzling. The consistency of our results
 with other altitude determinations is discuused in section 5 where
 we also bring in a new empirical '1/3 rule' to relate core/cone
 pulse widths and show the importance of the polar flux tube filling
 factors for these considerations. Conclusions and directions for
 future work which can exploit the full potential of this work are
 in section 6.

\section{The core-cone phase offset}

 In many pulsars, it is seen that the core component is shifted relative
 to the center of the conal pair. We call this offset in terms of arrival
 time, a 'lead' or a 'lag', depending on whether the core component is
 placed earlier or later in comparison with the conal centre. Both cases
 are seen (typical offsets $\simeq \pm 2^o$). As mentioned above the most
 direct explanation for these shifts is a combination of contributions
 from differential aberration and the {\it mfs}. At all
 altitudes, the contribution of toroidal component of the magnetic field
 of a pulsar works in a sense opposite to that of aberration. Aberrration
 tries to cause the arrival times to be earlier compared to the case when
 aberration is ignored. Similarly the {\it mfs} tends to delay the
 arrival. A consistent treatment should include both simultaneously.
 and accordingly we calculate here the net phase offset of a pulse
 component centre.

\subsection{The aberration offset}

 In KS we have considered the kinematical effects including aberration
 on pulsar beams in the polar cap model, and those fomulae we adapt to
 the present context. Using the same formalism and notation as in KS,
 we use two co-ordinate systems, one with the rotation axis as the
 $z$-axis (unprimed co-ordinates) and the other in which the magnetic
 axis is the $z^\prime$-axis (primed co-ordinates). The magnetic dipole
 lies in the $y-z$ plane at an angle $\alpha$ to the rotation axis.

 An emission point will have co-ordinates $(r, \theta_e, \varphi_e)$
 where $r$ is the emission altitude. The radiation is supposed to be
 emitted in a direction $(\theta_r, \varphi_r)$ tangent to the magnetic
 field lines. The latter is obtained by appropriately correcting
 ($\theta_e, \varphi_e$) for the tangential direction at any given
 point (the usual $\frac{3}{2}$ factor for small $\theta_e$). The
 polar cap boundary is defined by the locus of those angles
 ($\theta_{eb}, \varphi_{eb}$) which describe the feet of last open
 field lines. Therefore, the emission can emanate from the emission cap
 whose boundary is defined by the corresponding angles
 ($\theta_{rb},\varphi_{rb}$). As shown in KS the effects due to the
 stellar gravitational field namely
 change in the dipole geometry and light bending are mutually opposite
 and leave the Goldreich-Julian (1969) type of beam essentially unaltered.
 These effects rapidly become insignificant for altitudes beyond a few
 stellar radii. More importantly they affect the emission cap in a
 symmetrical way through an overall shrinking. We therefore ignore these
 general relativistic effects here. The directions ($\theta, \varphi$)
 after aberration are denoted by angles ($\hat \theta, \hat \varphi$).
 The aberrated values $(\hat \theta_r, \hat \varphi_r )$ of the angles
 $(\theta_r, \varphi_r)$ are given by (KS - Eqs. 37 and 38),

\begin{equation}
cos \, \hat{\theta_r} = \frac{cos \, \theta_{r}} {\gamma [1 + v \, sin \,
\theta_{r} \, sin \, (\varphi_{r} - \varphi_e)]},
\label{tab1}
\end{equation}
and
\begin{equation}
tan \, (\hat \varphi_r - \varphi_e) = \frac{\gamma [sin \,
\theta_{r} \, sin \, (\varphi_{r} - \varphi_e) + v]} {sin \,
\theta_{r} \, cos \, (\varphi_{r} - \varphi_e)}.
\label{pab1}
\end{equation}

\noindent where $\gamma= (1-v^2)^{-1/2}$ is the Lorentz factor and with

\begin{equation}
\label{vab1}
\xi =  r / r_L, \hspace{1cm} v  = \xi sin \theta_e
\end{equation}

 \noindent is the corotation velocity of the emission point at altitude
 $r$ and $r_L$ is the light cylinder radius.

 One can transform the $\hat \theta_r$ and $\hat \varphi_r$ values to
 the magnetic coordinates through

\begin{equation}
\label{r2m1}
 sin \, \hat \theta_r ^\prime \, cos \hat \varphi_r ^\prime = sin \, \hat
 \theta_r \, cos \, \hat \varphi_r,
\end{equation}

\begin{equation}
\label{r2m2}
 sin \, \hat \theta_r ^\prime \, sin \, \hat \varphi_r ^\prime = sin \,
 \hat \theta_r \, sin \, \hat \varphi_r \, cos \, \alpha - cos \, \hat
 \theta_r \, sin \, \alpha,
\end{equation}

\begin{equation}
\label{r2m3}
 cos \, \hat \theta_r ^\prime = cos \, \hat \theta_r \, cos \, \alpha
 + sin \, \hat \theta_r \, sin \, \hat \varphi_r \, sin \alpha .
\end{equation}

 The pulses will be observed by us when the direction
 $(\hat \theta_r, \,\hat \varphi_r)$ coincides with our line of sight.
 If the direction of line of sight is specified by angles
 $(\theta_{LS}, \, \varphi_{LS})$
 then the sweep of the line of sight through the emission cap is given
 by

\begin{equation}
\label{tpls}
 \hat \theta_r = \theta_{LS} \hspace{2.0cm} \hat \varphi_{r} =
\varphi_{LS}
\end{equation}

\noindent where the angle $\hat \varphi_{r}$ will change by an amount
 $\Omega \, t$ in time interval $t$ but $\hat \theta_r$ remains
 constant. Here $\Omega$ is the pulsar angular velocity.

 Values of $\hat \theta_r$ and $\hat \varphi_r$ are larger than those
 of the unaberrated angles $\theta_r$ and $\varphi_r$. For the special
 case of points lying on the meridian including the rotational and
 magnetic axes the formulae are simpler. In this case we confine
 attention to the $y-z$ plane, i.e., $\varphi_e = \varphi_r = \pi/2$.
 Consequently,
\begin{equation}
cos \, \hat{\theta_r} = \frac{cos \, \theta_{r}} {\gamma },
\label{tab2}
\end{equation}
 and
\begin{equation}
tan \, (\hat \varphi_r - \frac {\pi} {2}) = \frac{\gamma \, v}
 {sin \, \theta_{r}},
\label{pab2}
\end{equation}

 After including aberration the impact angle $\beta$ needs care
 in its definition. In absence of aberration the usual definition
 is $\beta = \theta_r - \alpha$ where $\varphi_r = \pi/2$ is
 understood. Now we can define it for example as
 $\beta = \hat \theta_r - \alpha$ with $\hat \varphi_r = \pi/2$
 understood. The relation of $\beta$ with the
 locations $(\hat \theta_{r}, \hat \varphi_{r})$ is then not
 simple. With the desired definition of $\beta$, its
 relation with $\hat \theta_{r}$ and $\hat \varphi_{r}$ can
 be derived using Eqs. \ref{tab1} and \ref{pab1} or Eqs. \ref{tab2}
 and \ref{pab2}.

 Henceforth, we consider the longitude offset only for the simple
 case of the direction of the magnetic axis. Firstly this should
 suffice for understanding the gross properties of the offsets
 which is our aim here, and secondly because the corresponding
 {\it mfs} offsets are available only for this case (Shitov 1983).
 Therefore, in addition to $\varphi_e = \varphi_r = \pi/2$ we
 also have $\theta_e = \theta_r = \alpha$. Now
 $v = \xi \, sin \, \alpha$ and then

\begin{equation}
 cos \, \hat{\theta_r} = \frac{cos \, \alpha} {\gamma },
\label{tab3}
\end{equation}

\begin{equation}
 tan \, (\hat {\varphi_r} - \frac {\pi}{2}) = \gamma \, \xi,
\label{pab3}
\end{equation}

 Eqs. \ref{pab2} and \ref{pab3}, in the nonrelativistic limit reduce
 to the often used $v/c$ formula.

 The offset in longitude introduced by aberration is
\begin{equation}
 \varphi_{ab} = \hat {\varphi_r} - \varphi_e = \hat {\varphi_r}
 - \frac {\pi}{2} = \gamma \, \xi,
\label{dpab}
\end{equation}

\subsection{The magnetic field line sweepback and the net offset}

 Equating the pulsar braking torque to the product of stellar magnetic
 moment and the toroidal component of the magnetic field caused by the
 stellar magnetic dipole radiation Shitov (1983, 1985) derived the
 toroidal magnetic field component $B_t$ as

\begin{equation}
 B_t =  \frac {2 \sqrt{\pi}} {3} B_0 \, (R_*/r_L)^3 \, sin^2 \alpha
\label{sh1}
\end {equation}

 where $B_0$ is the polar surface magnetic field. The magnitude
 of the magnetic field at a distance $r$ is $B(r)$.
 Presence of $B_t$ changes the azimuth of the co-rotating and radiating
 charges by an amount $\varphi_{mfs}$ such that

\begin{equation}
 tan \, \varphi_{mfs} = \frac {B_t} {B(r)} = \frac {r^3 B_t} {R_*^3 B_0} =
 \frac {2 \sqrt{\pi}} {3} \xi^3 sin^2 \alpha
\label{sh2}
\end {equation}
 which depends on $r = \xi \, r_L$ but not on $B_0$. (Shitov quotes and
 uses this in the small angle approximation).

 Eq. \ref{sh1} was derived with various simplifying assumptions. Here
 we use this result without any
 modifications. It should be kept in mind that a proper derivation of
 this equation should make use of the magnetic fields given by the
 full Deutsch solution (Deutsch 1955). In addition the toroidal
 component will also get a contribution from the stellar wind current
 which though unknown at present, still may be contributing to the
 pulsar slowdown (Goldreich and Julian 1969). Both these factors
 would warrant a much more involved analysis which is beyond the
 scope of this work. We intend to address these later. We also use
 Eq. \ref{sh2} as if it were valid for all emission altitudes.
 Actually we expect that a more precise handling of the {\it mfs}
 would certainly modify or even radically alter this equation for
 high altitudes. Therefore caution is advised in using Eq. \ref{sh2}
 for $\xi \simeq 1$. Similarly, it would be proper to correct the
 emission direction due to the {\it mfs} first and then aberrate it.
 This would give us the net phase offset $\Delta \varphi$. Due to the
 crude way in which we are including the {\it mfs} we here simply
 subtract the two contributions. This would be certainly adequate for
 low emission altitudes where both contributions are small. For higher
 altitudes we are assuming that the offsets though not precise will
 still be qualitatively correct. Thus, for any emission component,

\begin{equation}
 \Delta  \varphi(\xi) =  \varphi_{ab} -  \varphi_{mfs}
 \label{dp}
\end{equation}
 where $\varphi_{ab} = \hat \varphi_r - \varphi_e$ is given by Eq.
 \ref{pab3} and $\varphi_{mfs}$ by Eq. \ref{sh2}.

 We vary the emission point ($\xi$) from the stellar surface to the
 light cylinder along the radial direction. As mentioned above we
 look at $\Delta \varphi$ for the magnetic axis direction. It should
 be kept in mind that this gives a somewhat distorted picture of
 things because approximating the offset by
 that for the magnetic axis ignores the differential aberration over the
 emission cap which may be substantial at high altitudes.

\begin{figure}[t]
\epsfig{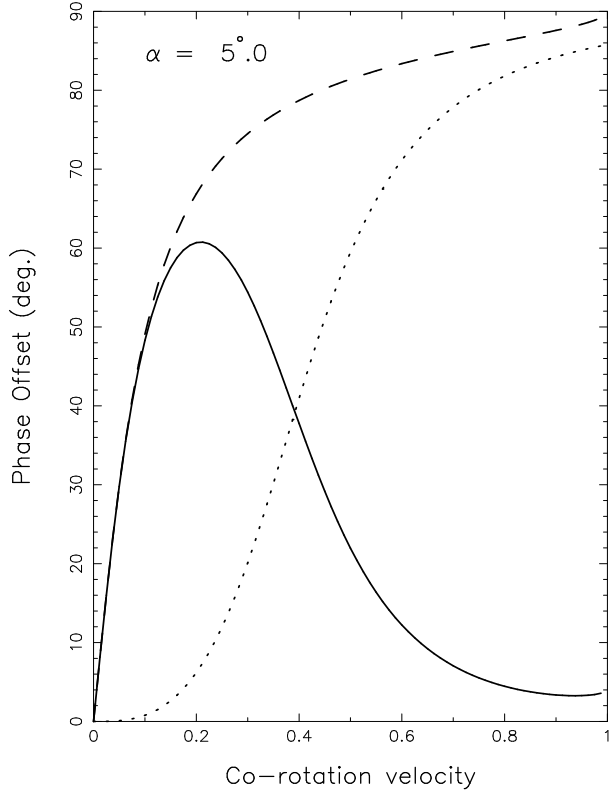}
\caption{Phase offsets of the emission point as a function of
 corotation velocity for $\alpha = 5^\circ$. Dashed line is the
 contribution $\varphi_{ab}$ due to aberration and dotted line is
 $\varphi_{mfs}$ due to {\it mfs}. Solid line is the net offset
 $\Delta\varphi(\xi)$ of Eq. \ref{dp}.}
\label{f1}
\end{figure}

 Note that in the limit in which light cylinder is approached, i.e., as
 $\xi \, sin \, \alpha \rightarrow 1$,
\begin{equation}
 tan \, \varphi_{ab} \rightarrow \infty, \hspace{1.0cm}
 i.e., \hspace{0.5cm} \varphi_{ab} \rightarrow \pi/2,
\label{lcab}
\end{equation}
 and,
\begin{equation}
 tan \,  \varphi_{mfs} \rightarrow 1.18 \, sin^2 \alpha.
\label{lcmfs}
\end{equation}

 If $\alpha \simeq \pi/2, \,  \varphi_{mfs} \simeq 1 \,rad$ and the net
 phase shift $\Delta \varphi$ would be $\simeq 0.57 \,rad$.

 As is customary, we assume that narrow-band emission emanates from a
 given altitude. With $r_{core}$ and $r_{cone}$ as altitudes for core
 and conal components respectively, we denote the phase offset between
 the core peak and the conal center  by $\Delta_{cc}$ and it is given by,
\begin{equation}
 \Delta_{cc} = \Delta  \varphi(\xi_{core}) -  \Delta  \varphi(\xi_{cone}).
 \label{dcc}
\end{equation}

 In addition, we also define
\begin{equation}
 \Delta r = |r_{core} -  r_{cone}|,
 \label{dr}
\end{equation}
\begin{equation}
 \Delta v = |\xi_{core} -  \xi_{cone}| \; sin \, \alpha .
 \label{dv}
\end{equation}

\begin{figure}[t]
\epsfig{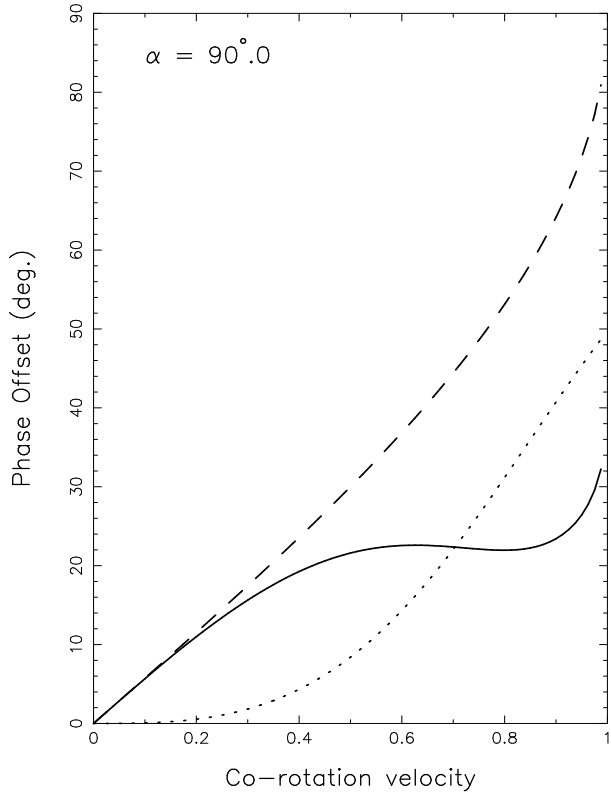}
\caption{Same as Fig. \ref{f1} but for $\alpha = 90^\circ$.}
\label{f2}
\end{figure}
\section{Calculated offsets}

 The latitude offsets (i.e., $\hat{\theta_r} -\theta_r$) though
 interesting can not be directly related
 to observations presently available. We therefore concentrate only on
 the longitude offsets $\Delta  \varphi$. Using Eq. \ref{dp} we have
 calculated the offsets $\Delta \varphi$ over the range of co-rotation
 velocity for various values of $\alpha$. Figs. 1-4 show these
 plots for $\alpha$ values as labelled.

\begin{figure}[t]
\epsfig{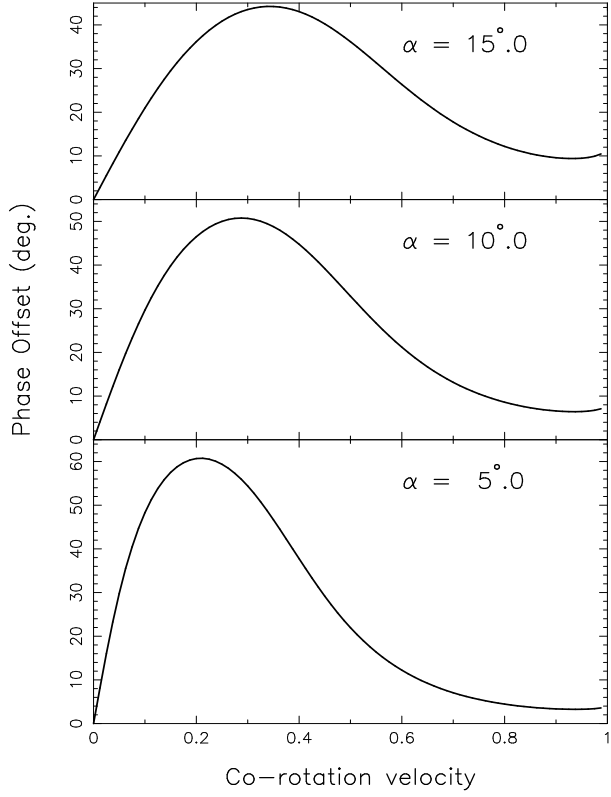}
\caption{Net offsets $\Delta\varphi$ vs. co-rotation velocity for
 various $\alpha$ as labelled.}
\label{f3}
\end{figure}

\begin{figure}[t]
\epsfig{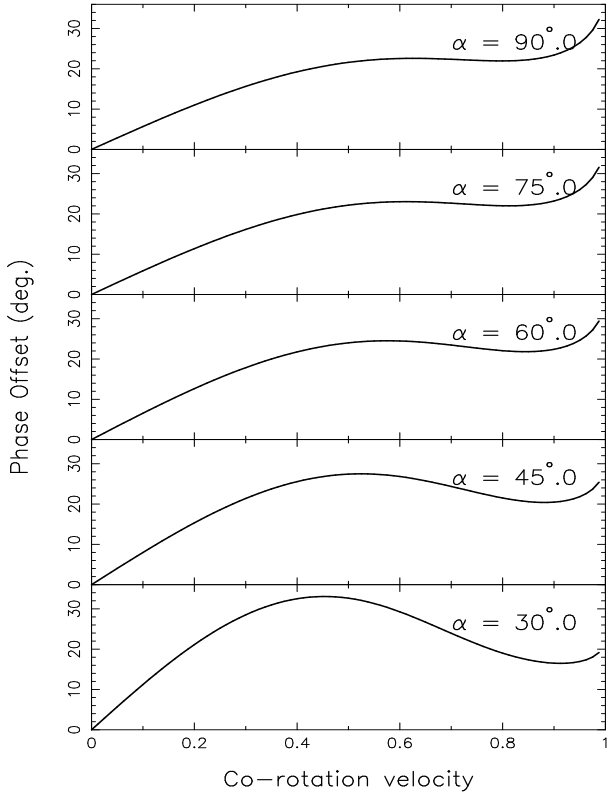}
\caption{Same as Fig. \ref{f3} but for different $\alpha$ values.}
\label{f4}
\end{figure}

 Figs. \ref{f1} and \ref{f2} show the individual contributions
 due to aberration and {\it mfs} and the net offsets in the small
 and large $\alpha$ cases of $5^\circ$ and $90^\circ$, while Figs.
 \ref{f3} and \ref{f4} show the net offsets, $\Delta \varphi$, as a
 function of the co-rotation velocity of the emission point.
 Qualitatively, the plots share many common features.

 Most notably the net offset is positive for all values of $\alpha$ and
 co-rotation velocity $v$. \emph {This means that aberration is always
 winning over \emph {mfs} at all altitudes and inclination angles}
 as is evident from Figs. \ref{f3} and \ref{f4}. The \emph{mfs}
 contributes insignificantly untill $v$ increases to $0.2$.
 The magnitude of
 {\it mfs} relative to aberration picks up for intermediate values of $v$,
 and the net offset is seen to vary in an interesting manner. For all
 $\alpha$, $\Delta  \varphi$ has a peak whose location shifts to higher
 values of $v$ as $\alpha$ increases. The picture is more or less the same
 for all inclination angles above about $20^\circ$. The phase shift,
 $\Delta  \varphi$
 first rises upto about half the distance to the light cylinder (we call
 this part of the curve as the rising branch) and then slopes down, rising
 again at $v$ around $0.8 \,c$. \emph {The peak value of $\Delta  \varphi$
 can be as large as $60^\circ$ for $\alpha = 5^\circ$, decreasing to
 $\cong 30^\circ$ for $\alpha = 45^\circ$}. For higher $\alpha$ the
 maximum is attained at $v =1$.

 We see that for a given offset, the emission altitude is not uniquely
 determined by $\Delta \varphi$. For $\alpha \le 60^\circ$, the same
 $\Delta \varphi$ is possible for two altitudes, while for
 $\alpha \ge 60^\circ$, a whole interval of altitudes is possible. Thus
 the altitude is a multivalued function of the offset $\Delta \varphi$.

 It is worthwhile to keep in mind that our plots when applied to a
 specific pulsar select a particular range of the $x$-axis. This range
 is determined by both the period of the pulsar and the assumed emission
 altitude. Clearly, for a given emission altitude, the co-rotation
 velocity will increase inversely with the pulsar period.

 Now we turn to the comparison of $\Delta_{cc}$ with observations.

\section{Comparison with observations}

 The core/cone phase offsets can constrain only the difference
 between the two emission altitudes. Therefore, it is not possible
 to make a direct comparison of 

\begin{table*}[b]
\begin{center}
\caption{Phase offsets for some pulsars}
\begin{tabular}{rclrcccll} \hline
  &           &     &               &      &        &         &         &
\\ \hline
SN& Pulsar    &Type &$\Delta \, _{cc}$&P   &$\alpha$&$\beta$  &$\Delta
r$&$\Delta v$ \\
  &	        &     &(deg.)         & (s)  & (deg.) &  (deg.) &$R_*$
& \\
\hline \\

 1&  0329+54  &T    &-2.0$^a$       &0.715 &30.8    &+2.9& 120&0.017 \\
 2&  0450-18  &T    & 0.0$^b$       &0.549 &31.0    &+4.9& ---& ---  \\
 3&  0523+11  &T    &+1.0$^k$       &0.354 &49.1    &+4.6&  30&0.015 \\
 4&  1451-68  &T    & 0.0$^b$       &0.263 &23.5    &+3.3& ---& ---  \\
 5&  1700-32  &T    & 0.0$^b$       &1.212 &47.0    &+1.7& ---& ---  \\
 6&  1804-08  &T    & 0.0$^c$       &0.164 &46.7    &+2.0& ---& ---  \\
 7&  1821+05  &T    &-1.9$^d$       &0.753 &32.0    &-1.5& 118&0.018 \\
 8&  1826-17  &T    &-1.5$^c$       &0.307 &39.6    &+0.7&  40&0.018 \\
 9&  1913+16  &T    &-12.0$^l$      &0.059 &46.0    &+1.0&  60&0.160 \\
10&  1917+00  &T    &+0.6$^e$       &1.27  &78.2    &-1.2&  62&0.010 \\
11&  1926+18  &T    &+0.6$^d$       &1.221 &35.0    &  - &  66&0.006 \\
12&  1933+16  &T    &-0.5$^f$       &0.359 &64.9    &+1.0&  16&0.008 \\
13&  1946+35  &T    & 0.0$^b$       &0.717 &43.1    &+4.4& ---& ---  \\
14&  2045-16  &T    &-2.1$^b$       &1.96  &37.0    &+1.1& 330&0.022 \\
15&  2111+46  &T    &-3.6$^g$       &1.015 & 8.6    &+1.3& 320&0.010 \\
16&  2319+60  &T    & 0.0$^b$       &2.256 &19.0    &+2.3& ---& ---  \\
17&  1237+25  &M    &-1.3$^h$       &1.382 &53.0    &-0.9& 200&0.025 \\
18&  1737+13  &M    &-0.8$^i$       &0.803 &41.0    &-2.5&  52&0.009 \\
19&  1857-26  &M    &-1.6$^b$       &0.612 &21.0    &-1.6&  81&0.010 \\
20&  2028+22  &M/T  &+1.9$^j$       &0.631 &50.0    &+5.5& 105&0.025 \\
\hline
\end{tabular}

References from where the $\Delta _{cc}$ value is obtained:\\
a - Lyne et al. (1971); b - LM88; c - Seiradakis et al. (1995);\\
d - Weisberg et al. (1986); e - RSW89; f - Sieber et al. (1975);\\
g - Kuzmin et al. (1998); h - R1; i - Blaskiewicz et al. (1990);\\
j - Hankins and Rickett (1986); k - Weisberg et al. (1999);\\
l - Cordes et al. (1990).
\end{center}
\end{table*}
 our results with observations so as
 to fit for individual pulsar parameters. This as will be seen would
 require that at least one of the core or conal emission altitudes is
 known. Still our results shed light on some important aspects of
 emission altitudes. Consequently, we focus on {\it trends} shown in
 observations rather than very precise agreements between them and
 the theory. For this purpose, although good and recent data is available
 with the European Pulsar Network maintained by MPI, Bonn (http://
 www.mpifr-bonn.mpg.de/div/pulsar/data/) we have used older pulse
 profiles which should suffice for our purpose. In our illustrative
 search we restricted to profiles from which offsets could be obtained
 by simple visual inspection, i.e., we typically searched for offsets
 among pulsars which are conal triples. We collected 59 triple (T) and
 multiple (M) profile pulsars (Lyne and Manchester 1988, henceforth
 LM88 and  R4) where the star types are as defined in R4. Of these, useful
 profiles could be identified for 20 pulsars. The pulsars which show a
 leading core component are 4 in number, 6 show zero phase offset and
 10 show a lagging core. (If M types are not included then these numbers
 are respectively 3, 6, and 7 out of a total of 16 stars of type T). Table
 1 shows these pulsars and their characteristics. The phase offset values
 generally correspond to 400 MHz data from the references listed there.
 The star types, $\alpha$ and $\beta$ values, here as well as in Table 2
 which comes later, are from R4 or LM88. They no doubt
 will have some inter-relation with $\Delta_{cc}$. Within the scope of
 this paper, we do not dwell on these details.

 Although very large altitudes may come into play for emission from
 outer gaps (see Hirotani (2000) and references therein), from all
 indications so far it seems that usual pulsar
 emission altitudes are $< 0.2 \, r_L$. As noted before our inclusion of
 {\it mfs} also has limitations for very large altitudes. Therefore
 to avoid needless complexity we first concentrate on the rising
 branch of the $\Delta\varphi$ curves in Figs. 3 and 4.

 To derive a core/cone altitude difference all one needs is to assume
 appropriate altitudes for core and conal components, read off  the
 corresponding offsets from the curve and compare their difference
 with the observed values. Values of $\Delta r$ and $\Delta v$
 derived in this manner are included in Tables 1 and 2. Based on
 such a view of altitudes, explanation of offsets in Table 1 does
 not present any problem. In fact, the offsets probably are pointing
 towards some properties of emission altitudes.

\subsection{What do signs of observed offsets indicate ?}

 Assume that the core emission comes from the stellar surface as the
 simplest interpretation of the observed core widths (R4) : then, the
 corresponding $\Delta  \varphi$ for core is very small and almost same
 as that for $v = 0$. The conal emission must now come from a higher
 altitude (e.g., $10 - 20 \, R_*$ according to R6). It is obvious from
 the figures that if the offsets are attributed to the effects
 considered here then the core component can only lag. Observationally
 both cases are seen in Table 1. A majority of stars do show a lag, but
 the stars which do not conform to this are not negligible ( about
 $50 \%$). Our sample can not reliably tell what will be
 the actual fraction of such stars. But in all such cases clearly the
 core emission can not come from the surface. {\it The signs of observed
 offsets thus indicate that each of the possibilities, i.e.,
 $r_{core} <, =, or > r_{cone}$ can occur}.

\subsection{What do magnitudes of observed offsets indicate ?}

 The magnitude of $\Delta_{cc}$ in Table 1 generally is $\cong 2^\circ$.
 The first notable thing is its smallness compared to the maximum
 possible values seen in the Figs. 3 and 4 ($\cong 30^\circ - 60^\circ$
 depending on $\alpha$ values). It is necessary to stress here that a
 small value of $\Delta_{cc}$ which implies a small $\Delta v$
 will translate to a small $\Delta r$ only for some values of
 $\alpha$ and $P$, even when altitudes lie on the rising branch.
 \emph {The magnitudes of $\Delta_{cc}$ thus indicate that $r_{core}$
 and $r_{cone}$ are closeby in comparison to $r_L$ but not necessarily
 so in terms of $R_*$}.

 As examples let us look at altitude differences for two pulsars in which
 the core lags. Using Eq. \ref{dcc} for $\Delta_{cc}$ we get an altitude
 difference for PSR 0329+54 as $\Delta r \cong 115 R_*$, and for
 PSR 1913+16 as $\cong 65 R_*$. We shall return in Section 5 to these
 values of $\Delta r$ for lagging cores and the zero offset cases to see
 if they can be reconciled with altitudes derived in HX, R4 and R6.
 If on the other hand we consider two examples from pulsars in which
 the core leads, i.e., PSR 1917+00 and PSR 2028+22, then we get $\Delta r$
 as $\cong 60 R_*$ and $\cong 100 R_*$ respectively. In these pulsars the
 core altitude must be larger than the conal one, which is not possible
 with altitudes given in HX, R4 or R6.

\subsection{Can both core and cone emisison altitudes be very large ?}

 Although we are restricting to emission altitudes $< 0.2 \, r_L$, if
 we consider altitudes $> 0.2 \, r_L$ a small $\Delta_{cc}$ is still
 possible. The curves in Figs. \ref{f3} and \ref{f4} allow, in principle,
 widely different altitudes which lie on the opposite sides of the peak
 for smaller $\alpha$. Though widely different, these altitudes
 correspond to about the same small $\Delta  \varphi$ leading to even
 smaller $\Delta_{cc}$. For larger values of $\alpha$ the peak in
 $\Delta \varphi$ is not pronounced so a small value of $\Delta_{cc}$
 can occur for a larger range of $\Delta r$ and the correlation
 that a larger lead corresponds to a larger altitude would not hold.
 As remarked earlier, our results warrant caution for high altitudes
 beyond the peak. There is one more caveat. For $\xi > \frac {2}{3}$ the
 angular size of the emission cap exceeds $180^\circ$, and emission
 therefore must come from only a part of the cap. Without some further
 insights there is not much advantage in considering very large emission
 altitudes for both core and conal components especially when
 $\Delta_{cc}$ is small. This notwithstanding, for larger values of
 $\Delta_{cc}$ at least one of the altitudes must be very large. This
 case will be characterised by the core/cone altitudes being widely
 different. We discuss this in the next subsection.

\subsection{Can core and cone emisison altitudes be widely different ? :
 Peculiar data}

 As for the very large values of $\Delta  \varphi$ which are seen in our
 plots, do they really manifest in some pulsars? If so then some
 interesting possibilities emerge.

 One possibility is that a large offset makes the core component merge
 with one of the conal ones. The core then would seem absent or barely
 discernible. The conal pair in that case will not display symmetrical
 intensity and polarization, and typical spectral properties. In some
 triples, i.e., of type T, this may be a cause of confusion in their
 classification. One would expect this to be likely in those pulsars
 which have small $\beta$ and absent or weak core component with conal
 components showing confusing characteristics. Observationally such a
 possibility was noted long ago but was not pursued as is demonstrated
 by the following quote,'...we must recognize the possibility of
 double profiles which are actually partially merged triples -
 particularly where there is a marked lack of symmetry between the
 two components in amplitude, width, or polarization behavior..'
 (R1 p. 343).

 Another interesting possibility is that the offset is so large that the
 core may even be displaced outside the conal pair.

 A search through observed profiles with this in mind led us to the
 stars listed in the Table 2. These cases are of great interest and may
 throw much light on pulsar emission altitudes. The longitude offsets
 quoted in the Table 2 as for Table 1 were read from published profiles
 and  though not very precise, suffice for our purpose. For first eight
 stars, the $\Delta_{cc}$ values equal one-half of the longitude
 separation between the two visible peaks, and are upper limits because
 the core/cone distinction is not completely clear in the profile. The
 star types are again as in R4.

\begin{table}[h]
\begin{center}
\caption{Large phase offsets cases}
\begin{tabular}{lcllcccll} \hline
  &            &     &                 &      &        &         &
& \\ \hline
SN& Pulsar     &Type &$\Delta \, _{cc}$&P     &$\alpha$&$\beta$  &$\Delta
r$&$\Delta v$ \\
  &	         &     &(deg.)           & (s)  & (deg.) &  (deg.) &$R_*$
& \\
\hline \\
 1 &  1802+03  &---            &$\le+4.1$ &0.219 &---- &----&  75&$\le
0.07$ \\
 2 &  1822-09  &T$_{1/2}$      &$\le-7.5$ &0.789 &86.0 &+1.1& 490&0.130 \\
 3 &  1842+14  &T/S$_t$        &$\le+3.9$ &0.375 &63.0 &+4.2& 125&0.063 \\
 4 &  1859+03  &S$_t$/T$_{1/2}$&$\le-5.6$ &0.655 &38.7 &+3.6& 310&0.063 \\
 5 &  1907+10  &S$_t$          &$\le-3.8$ &0.284 &90.0 &----&  91&0.067 \\
 6 &  1920+21  &T              &$\le+4.5$ &1.078 &47.0 &+1.2& 420&0.060 \\
 7 &  2020+28  &T              &$\le+5$   &0.343 &71.0 &+3.6& 145&0.080 \\
 8 &  2224+65  &T$_{1/2}$      &$\le+15$  &0.683 &16.0 &+3.4& 900&0.080 \\
   &            &               &          &      &     &    &    &
\\
 9 &  0531+21  &T$_{1/2}$      &$+22$     &0.033 &90.0 &----& 100&0.650 \\
10 &  0823+26  &T$_{1/2}$      &$+30$     &0.531 &90.0 &----&2480&0.980 \\
11 &  1055-52  &T              &$+16$     &0.197 &90.0 & 0.0& 300&0.330 \\
12 &  1742-30  &T              &$-9$      &0.367 &57.0 &----& 290&0.140 \\
\hline
\end{tabular}
\end{center}
\end{table}

 The first eight stars in Table 2 illustrate the merging of the core
 component with one of the conal ones. The last four correspond to the
 core components being displaced outside the cone. We discuss them one
 by one.

 1. PSR 1802+03 :

 "The extremely high linear polarization and flat position angle of the
 leading component argue that it is a core or possibly a merged core and
 cone, with the trailing component being conal" (Weisberg et al. 1999).
 The core thus leads by $\le 4.1^\circ$.

 2. PSR 1822-09 :

 This pulsar has been described variously as "most interesting" and
 "intriguing" (Manchester et al. 1980, Rankin 1986, hereafter R3) and
 is identified as a one-sided triple (T$_{1/2}$). Its profile strikingly
 resembles that of the Crab pulsar (both exhibit interpulses). The main
 pulse profile is double with the leading component ("precursor") showing
 conal characteristics. The core component in this star is identified as
 the leading component of the composite second "component" of the main
 pulse (R6). From a 1612 MHz profile (Manchester et al. 1980; see also
 R3), we read the core component lag as $7.5^\circ$.

 3. PSR 1842+14 :

 Similar to PSR 1802+03 in 1 above (Weisberg et al. 1999) with core lead
 $\le 3.9^\circ$.

 4. PSR 1859+03 :

 Profile is similar to PSRs 1802+03 and 1842+14 but core is identified
 here from its circular polarization signature (Weisberg et al. 1999),
 and lags by $\le 5.6^\circ$.

 5. PSR 1907+10 :

 At 21 cm and higher frequencies, the leading component is apparently
 conal while the primary component is a core, possibly with another
 weak one merged on to its trailing edge (Weisberg et al. 1999). The
 core lag is $\le 3.8^\circ$.

 6. PSR 1920+21 :

 In this star the core component is so early that it virtually overlies
 the leading conal outrider (R6; see also Rankin et al. 1989, hereafter
 RSW89). A 1418 MHz profile (Weisberg et al. 1999) gives the core lead
 as $4.5^\circ$.

 7. PSR 2020+28

 In the profile of this pulsar as seen in Sieber et al (1975) and Cordes
 et al (1978), the two conal components obscure the weak core
 component on the trailing edge of the leading component (R6) which
 also has been termed as a bridge (Weisberg et al. 1999). The profile
 can be modelled as the sum of three Gaussian components, in which
 the steep spectral index above 800 MHz of the bridge "suggests" that
 it is a core (Weisberg et al. 1999). From the profile at 430 MHz in
 Sieber et al. (1975), we find that the core leads by $5^\circ$.

 8. PSR 2224+65 :

 This pulsar is classified as a partial cone star in LM88, while according
 to R6 the profile has two well separated components of which the first
 seems to be the core component (steeper spectrum), which overlies the
 leading conal outrider. From a 1420 MHz profile in Seiradakis et al.
 (1995), we find that the core leads by $15^\circ$.

 The next four pulsars show cores with offsets so large that they have
 been displaced outside the conal pair.

 9. PSR 0531+21 :

 The Crab pulsar though very well studied is not well understood. The
 precursor, "by virtue of its softer spectrum .... should probably be
 regarded as a core component" (R4). The precursor-main pulse profile
 is interpreted as a triple profile in which  the leading conal
 outrider is missing (R6). Interpreted in this manner the core leads
 by $22^\circ$ ( 318 MHz profile in McCulloch et al. 1976). Due to the
 T$_{1/2}$ nature of the profile, $\Delta_{cc}$ can not be determined
 unambiguously. We have taken the core lead as of the same order as
 the separation $22^\circ$. See the similar case of PSR 1055-52
 discussed below.

 10. PSR 0823+26 :

 "The main pulse is readily identified as a core component (...
 circularly polarized signature), and the weak interpulse may be one
 also. The overall main pulse-postcursor complex might again be regarded
 as a T$_{1/2}$ profile  but the $30^\circ$ spacing is again rather large"
 (R4, see also RSW89). Again this seems to be a case where the core
 is leading by $30^\circ$ (the 1400 MHz profile in RSW89).

 Note the magnitude of $\Delta_{cc}$ in this pulsar. With $\alpha$ for
 this pulsar quoted as $90^\circ$ the core emission occurs very close
 to the light cylinder (see Eqs. \ref{lcab} and \ref{lcmfs}) if our
 {\it mfs} can be relied on. From our plots a $30^\circ$ offset can
 be obtained at a lower altitude for a small $\alpha$. Therefore,
 both, a better treatment of {\it mfs} and a reverification of the
 $\alpha$ value would be of help.

 11. PSR 1055-52 :

 The PSR 1055-52 shows great similarity to the Crab pulsar, though
 there are some minor differences. "Clearly we should regard the
 main pulse as having a T profile" (R4). The interpulse and the
 precursors in both of these very strongly display properties typical
 of core components (McCulloch et al. 1976, R1). Thus, like PSR
 0531+21 this pulsar is also an interesting example in which the
 core has been displaced outside the conal pair with a lead of
 $16^\circ$ (the 635 MHz profile in McCulloch et al. 1976).

 12. PSR 1742-30 :

 "The core component of this very interesting pulsar seems to follow
 the trailing 'outrider'- a very unusual configuration !" (R6). "In
 pulsar 1742-30 we then seem to have either a five-component (M)
 profile with missing components 4 and 5 or a triple (T) profile in
 which the core component lies outside of the conal outriders !"
 (Xilouris et al 1991; see also Seiradakis et al. 1995). This is a
 case in which the core lags; we read it as amounting to $9^\circ$
 from a 1700 MHz profile in Xilouris et al. (1991).

 \emph {Table 2 thus clearly demonstrates that widely different
 core/cone altitudes, differing by a significant fraction of $r_L$,
 do occur in some pulsars.}

 In the first three pulsars in this group, the nomenclature main
 pulse has been used in varied senses by different authors. But if the
 separation between similar components is considered the time lag comes
 much closer to half the pulsar period, in fact within the expected errors
 it may be exactly $180^\circ$. For the above pulsars the
 interpulse is core emission and therefore should be compared only with
 the core component of the non-interpulse. {\it  We believe that the
 mystery of main pulse - interpulse separations being much different
 from the expected value of $180^\circ$ may get completely solved
 if the data is reanalysed to obtain  more precise values of
 $\Delta_{cc}$ for these pulsars than what we have.} We hope to report
 on this soon.

\section{Consistency with other altitude determinations}

\subsection{Consistency with the core width relation}

 The relation derived from observations by Rankin (R4) for core component
 widths besides being a very good fit to observations also lends itself to
 a very natural interpretation that the core emission originates from the
 stellar surface and the full polar cap of a magnetic dipole. The polar
 cap (i.e., the emission cap on the stellar surface) size is the usual
 one for a dipole as given by Goldreich and Julian
 (1969). Recently we have analyzed this further after inclusion of general
 relativistic effects due to the stellar mass (Kapoor and Shukre 2001).
 The gravitational effects lead to constraints that $M_{pulsar} < 2.5 \,
 M_\odot$ and $R_{pulsar} \le 10.5 \, km$. None of the available equations
 of state of the neutron star matter can satisfy the radius constraint
 for a reasonable mass, leading one to the conclusion that pulsar are
 strange stars rather than neutron stars. From the present work it
 appears that core emission may not originate from the stellar surface.
 Is it possible to choose between these two opposite views of core
 altitudes ? Later in the section we show that assumptions needed for
 the usual interpretation of the Rankin core width relation need a
 revision, and therefore the constraints on pulsar radii would also
 get revised.

\subsection{Consistency with pulse component widths}

 The altitude difference derived using $\Delta_{cc}$ must be consistent
 with altitudes determined by other methods. Generally speaking this
 seems a difficult task, because the differences in core/cone altitudes
 derived using $\Delta_{cc}$ are much larger than the individual values
 derived  by other methods, see e.g., HX, R4 and R6. HX use the BCW method
 while R4 and R6 use pulse widths. As we discuss later the
 altitudes derived using the BCW offset would require a reanalysis of
 data for a meaningful comparison with $\Delta r$ derived here.
 Before this comparison a similar and refined reanalysis
 for deriving better $\Delta_{cc}$ values will also be needed.

 The altitudes derived from pulse widths in R4, R6
 however can be increased while maintaining the agreement with observed
 pulse widths. Since in the dipole geometry the cap radius $\rho$ scales
 along a given flux tube
 with the altitude $r$ as $\sqrt{r}$, an increased altitude can still
 lead to the same pulse width if one assumes that instead of the full
 emission caps, only parts of them give out radiation. This allows one
 to reconcile the altitudes based on pulse widths with our altitude
 difference.

\subsubsection{The 1/3 rule and the filling factors}

 For this purpose, we introduce a simple empirical relation between
 core and conal widths which should be looked
 upon as a rule of thumb rather than a precise formula. Taking triple
 profiles as typical indicators of widths of the core and two conal
 components in one pulse, we take each one to occupy 1/3 of the
 total pulse width. This 'one-third' rule which has an observational
 origin will be seen to be of great utility and can be formalized as
 follows.

 The core component is emitted from an altitude $r_{core}$ where the
 full emision cap has the Goldreich-Julian angular radius
 $\rho_{core}$. If only a central part of the cap is emitting we
 describe it with a filling factor $f_{core}$ in terms of the radius
 (or equivalently the pulse longitude), and it is the ratio of the
 radius of the part which emits to $\rho_{core}$. The width of the
 core component is
\begin{equation}
 W_{core} = 2\, f_{core}\, \rho_{core}.
 \label{wcor}
\end{equation}

 The conal component similarly comes from an altitude $r_{cone}$ and
 a cap of Goldreich-Julian radius $\rho_{cone}$, but has a hollow cone.
 Its filling factor is $f_{cone}$, the ratio of the difference of the
 outer and inner radii of the emission region to $\rho_{cone}$. The
 width of each conal component is

\begin{equation}
 W_{cone} = f_{cone}\, \rho_{cone}.
 \label{wcon}
\end{equation}

 Now the 1/3 rule tells us that
\begin{equation}
 W_{core} = W_{cone}
 \label{13rul}
\end{equation}
 and invoking the $\sqrt{r}$ scaling of $\rho$, we get the cone/core
 altitude ratio as
\begin{equation}
 r_{cone}/r_{core} = [\rho_{cone}/\rho_{core} ]^2 = 4\, f_{core}^2 /
 f_{cone}^2\\
\label{rat}
\end{equation}

 Maximum value of $f_{core}$ is 1 when the core emission occurs in the
 full cap. For $f_{cone}$ the maximum is 2/3 and then the conal
 emission uniformly occupies the annulus between radii $\rho_{cone}/3$
 and $\rho_{cone}$. These maximum filling factors give $r_{cone}/r_{core}
 = 9$. The observed pulse widths will of course be usually less than $W$
 above due to the impact angle $\beta$ not being zero. To stay independent
 of the fortuitous value of $\beta$ which obtains for any observed pulsar,
 we base our discussion on $\rho$ values alone.

 \emph {Introduction of filling factors may appear as an unwarranted
 additional complication, but as Eq. \ref{rat} shows, they cannot be
 separated from the discussion of emission altitudes}. For example the
 work by Rankin (R4, R6) tacitly assumes that $f_{core}$ and
 $f_{cone}$ have their maximum values of 1 and 2/3 respectively. This
 by the 1/3 rule would give $r_{core}/r_{cone} = 1/9$. Rankin's
 analysis is based on pulse widths and is more refined than '1/3'
 thumb rule, and gives $r_{core} = R_*$ and $r_{cone} \cong 10 R_*$
 (We ignore the further subtleties of inner and outer cones).
 Therefore the 1/3 rule captures in a simple manner the core-cone
 width relation and at the same time brings into focus the filling
 factors.

 The altitudes derived by HX make use
 of the BCW offset and give $r_{core} \cong 21 \pm 18 \, R_*$ and
 $r_{cone} \cong 42 \pm 20 \, R_*$. These altitudes are determined
 independent of any assumption about the filling factors, but
 in conjunction with the 1/3 rule (see Eq. \ref{rat}) imply that
 the filling factors do not have their maximum values, in fact the
 mean values of altitudes imply a ratio of filling factors
 $f_{core}/ f_{cone} = 1/ \sqrt{2}$.

 We now reconsider in turn pulsar examples in which the core component
 lags, has zero offset and leads respectively.

\subsubsection{Lags}
 Returning now to the pulsars PSR 0329+54 and PSR 1913+16 which
 show lags, for altitude differences using the $\Delta_{cc}$ we get
 respectively $\Delta r \cong 115 R_*$, and $\cong  65 R_*$, which
 cannot be accomodated by R4, R6 or HX.

 Since altitudes derived in R4 and R6 give correct observed pulse
 widths, it would be desirable to maintain this. At the same time
 our $\Delta r$ can also be reproduced in the following manner. If we
 assume a particular value for $f_{core}/ f_{cone}$ then Eq. \ref{rat}
 gives us the altitude ratio. Combining this with the $\Delta r$ it
 is possible to derive $r_{core}$ and $r_{cone}$ individually, with
 the understanding that the higher altitude pertains to the leading
 component.

 This way if $f_{core}/ f_{cone} = 3/2$ as in R4, R6 then
 $r_{core} = 14 R_*$ and $r_{cone} = 129 R_*$ for PSR 0329+54.
 Similarly for PSR 1913+16 we get $r_{core} = 7 R_*$ and $r_{cone}
 = 72 R_*$. In this way $\Delta_{cc}$ and R4, R6 can be reconciled.

 For PSR 0329+54 the altitude quoted in HX is negative while
 PSR 1913+16 does not occur in their list. Also the altitude ratio
 $r_{cone}/ r_{core}$ for the general altitudes quoted by them varies
 between 0.6 and 20. If we take the mean value of 2 leading to
 $f_{core}/ f_{cone} = 1/\sqrt{2}$, we can reconcile also
 $\Delta_{cc}$ and HX. The altitudes then are $r_{core} = 115 R_*$ and
 $r_{cone} = 230 R_*$ for PSR 0329+54 and $r_{core} = 63 R_*$ and
 $r_{cone} = 126 R_*$ for PSR 1913+16. However, we do not attach any
 significance to this agreement and discuss it further in the
 summary subsection later.

 At this point the intimate relation between emission altitudes and
 filling factors should be clear.

\subsubsection{Zero offsets}

 In these cases  both core and conal emissions occur at the same
 altitude. Then Eq. \ref{rat} implies that $2\, f_{core} = f_{cone}$
 and further because in this case $\rho_{core} = \rho_{cone}$,
 we have $f_{core} + f_{cone} = 1$ and we find that $2\, f_{core}
 = f_{cone} = 2/3$. Apart from the information that their difference
 is zero, however, the altitudes can not be fixed any further even
 by using the 1/3 rule. The $\Delta_{cc}$
 corresponding to altitude differences in R4, R6 ($\cong 10 \, R_*$)
 or HX ($\cong 20 \, R_*$) would be small enough to agree with the
 observed ones. Consequently, unless the data is extremely precise
 the ratio $f_{core}/ f_{cone}$ is not determined, e. g., no choice
 can be made among either $3/2$, $1/\sqrt{2}$ or $1/2$.

 For zero offset cases when $\alpha \cong 30^{\circ} - 50^{\circ}$
 range, due to the multivaluedness of $r$ an infinite number of pairs
 of $r_{core}$ and $r_{cone}$ give the same $\Delta_{cc}$. But if
 either $r_{core}$ or $r_{cone}$ is found to be on the rising branch
 (say by using BCW) and such that its $\Delta \varphi$ is small enough
 to be possible only at one altitude (multivaluedness is avoided)
 then core and conal altitudes must lie close.

\subsubsection{Leads}

 Consider the pulsars PSR 1917+00 and PSR 2028+22 for which $\Delta_{cc}$
 respectively gives $\Delta r$ as $\cong 60 \, R_*$ and
 $\cong 100 \, R_*$. It is obvious that altitudes given by R4, R6 or HX
 are of no help here because they can not give the correct sign of
 $\Delta_{cc}$.

 We thus try to find $f_{core}/ f_{cone}$ in these cases using the 1/3
 rule. Assuming minimum posiible altitude for the conal component, i.e.,
 $1 \, R_*$, and using Eqs. \ref{wcor} and \ref{wcon} for these two
 pulsars we find $f_{core}/ f_{cone} = 1/16$ for PSR 1917+00 and $1/20$
 for PSR 2028+22. (The solid angle filling factors will be squares of
 these.) Other combinations consistent with $\Delta_{cc}$ values will
 give higher altitudes and  further reduce this ratio. \emph {Unlike
 the cases of lags or zero offsets here clearly the full open flux
 tube must have large regions which do not radiate at all.}

\subsubsection{Summary}

 In the above discussion of lags, leads and zero offset cases we
 have tried to reconcile the various altitude estimates provided
 the core/cone distinction is not ignored.

 Intrinsically the BCW method is an excellent method for altitude
 determination and enjoys the advantages that use of $\Delta_{cc}$
 also entails. In both BCW and HX a puzzling feature is that the
 derived altitudes have large uncertainties despite having quality
 data and improved physics. Also in some pulsars the BCW offsets come
 out with a wrong sign (BCW, HX). We feel that this may be a
 consequence of an insufficient disentangling of core and conal
 features. For example, in BCW the position angle curve is dominated
 by the conal wings while the intensity profile is dominated by the
 central core in some pulsars. The BCW method as also used by HX
 utilises the linear polarization position angle sweep, a typical
 conal property, despite the reference to pulsars as 'core-dominated'
 etc. The position angle sweeps for cores are well known to be
 usually not so clean. It seems that BCW offsets in effect lead to
 altitudes only of conal components but in cases where the core is
 or is not seen. Therefore, for core components using the circular
 polarization signatures may provide the way. Though there is some
 overlap in their and our pulsar lists, the reconciliation of
 altitude values quoted for specific pulsars by either BCW or HX
 with our $\Delta r$ here will not be meaningful; certainly so if
 we use $f_{core}/ f_{cone} = 1/\sqrt{2}$ which relates only to
 mean values of altitudes in BCW or HX. A case by case analysis
 can be done, but it may not prove of much value, because
 our $\Delta r$ can be determined precisely only when the core and
 conal components have been completely separated. Therefore, it is
 difficult to compare our $\Delta r$ with altitudes derived using
 the BCW method if the core/cone are not fully disentangled.

 The considerations using pulse widths tacitly need an assumption
 about the filling factors. What has emerged here is that the
 filling factors of core/conal emissions form an integral part of
 the discussion and an appropriate filling factor ratio can bring
 about an agreement between $\Delta r$ here and the altitudes
 derived using pulse width data. Since this has been brought about
 by considerations of $\Delta_{cc}$, we see that explanation of
 core/cone offsets by kinematic effects has implications for the
 emission mechanism. Below we show that some knowledge of these
 filling factors can be gained if we use the 1/3 rule, further
 demonstrating its utility.

\subsubsection{Variation of filling factors with altitudes}

 Though only the emission mechanism can provide the knowledge of
 filling factors it is possible to infer about one aspect of them
 from $\Delta_{cc}$. The variety in values of $\Delta_{cc}$ most
 probably reflects different values of $P$, $\alpha$ and $\beta$ for
 different pulsars. Imagine now pulsars arranged in an order such that
 successively $\Delta_{cc}$ decreases starting from a zero value. For
 the zero offset case we have $r_{core}=r_{cone}$, $2\, f_{core} =
 f_{cone} = 2/3$. Thus  $r_{cone}/r_{core} = 1$ and $f_{cone}/f_{core}
 = 2$. Now let the cone move up relative to the core. Combined with
 the 1/3 rule we see that this is equivalent to effectively keeping
 $f_{cone} = 2/3$ and increasing $f_{core}$ to a value $> 1/3$. Thus
 as $\Delta_{cc}$ becomes more negative the ratio $f_{cone}/f_{core}$
 decreases as $r_{cone}/r_{core}$ increases. Similarly we now
 imagine a progression of pulsars starting from zero offset and
 increasing $\Delta_{cc}$. In this case same reasoning tells us that
 cone is moving down relative to the core and at the same time the
 ratio $f_{cone}/f_{core}$ is increasing from a starting value of 2
 as $r_{cone}/r_{core}$ decreases. Thus an important conclusion follows
 : \emph {qualitatively we find that the ratio $f_{cone}/f_{core}$ is
 a monotoniaclly decreasing function of the altitude ratio
 $r_{cone}/r_{core}$}. It should be noted that this filling factor
 variation with altitude would hold at all emission frequencies.

\section{Conclusions}

 By attributing the longitude offsets between the centers of core and
 conal components to kinematical effects due to their having different
 altitudes of emission leads to the following :

 \begin{itemize}
 \item The combined offset due to aberration and \emph {mfs} always
 advances the arrival time for all altitudes and inclination angles
 $\alpha$.

 \item Core emission does not necessarily come from the stellar surface.

 \item Core emission altitudes, $r_{core}$ may be smaller, larger than or
 same as $r_{cone}$, the conal ones.

 \item For most pulsars $|r_{core} - r_{cone}|$ is small compared to
 $r_L$ but not necessarily compared to $R_*$. Indeed
 $|r_{core} - r_{cone}|$ is usually much larger than the individual
 altitudes ascribed to core and conal components so far.

 \item For some pulsars $|r_{core} - r_{cone}|$ could be comparable to
 $r_L$. This may resolve the mystery of main pulse-interpulse separations
 being much different from the expected $180^\circ$.

 \item Both core and conal emissions do not come from the full available
 part of the polar flux tube and their filling factors vary with the
 altitudes.
 \end{itemize}

 We do not wish to be more quantitative about the emission altitudes at
 this stage for two reasons. Firstly, on account of the previously
 mentioned limitations of the {\it mfs} formula we need a refinement
 in it. This should treat the sweepback more precisely using the
 Deutsch solution for the magnetic field of an oblique rotator and we
 intend to report on it soon. Secondly, the offset translates only into a
 difference between emission altitudes of the two components. It can
 determine individual altitudes only if at least one of the altitudes
 (core or cone) is known definitely.

 In the foregoing we have ignored the possible contributions to phase
 offsets by the emission mechanism(s). Can emission mechanisms
 contribute significantly to the phase offset of a pulse component? Note
 firstly that aberration and the 'mfs' are first order effects. As seen
 in Figs. 1 and 2, these offsets are individually quite large over most
 of the range of $v$ as well as $\alpha$. Similarly from Figs. 3 and 4,
 the same is true even when we deal with their difference, i.e., $\Delta
 \varphi$. Even at small (but not extremely small) values of $v$, and,
 also $\alpha$, the net offset is significant. For the emission mechanism
 to affect this, its contribution should also be equally significant. If
 this is true then our considerations would need modifications.

 On the other hand, irrespective of these dynamical contributions which
 are also not known, the effects considered by us must be present. From
 the foregoing there is ample indication that they suffice for describing
 the observed core/cone longitude offsets coming into play due to the
 emission mechanism selecting different altitudes of emission, perhaps
 strongly varying with period, inclination angle etc. In addition,
 they are providing constraints on the emission mechanisms, e.g.,
 the unexpectedly small filling factors for the core components for
 those pulsars in which core leads.

 Now, we briefly comment on multifrequency observations. As the
 frequency of observation decreases, $\Delta_{cc}$ remains negative
 but increases in magnitude for some pulsars (e.g., PSR 0329+54,
 see, e.g., Malov and Suleimanova 1998) while it decreases to such
 an extent that it changes from negative to positive for PSR 1821+05
 (see, e.g., Weisberg et al. 1999). Since our considerations bring
 in filling factor variations, both frequency and geometry of
 emission change with the altitude. Thus, our picture potentially
 can explain the otherwise perplexing observations and its
 application will lead to very insightful constraints on the emission
 mechanism. In addition, \emph { a rfm opposite of the
 conventional one is also a possibility}. This seems to have been
 observed by Rankin (2001, private communication).

 The full predictive potential of our framework can be realized if it
 is combined with a detailed analysis of morphological, polarization
 and multifrequency observations. We offer one concrete suggestion in
 this connection.

 The total intensity profile of a pulsar could be separated into
 individual Gaussian components. The circular polarization signature
 could be used to identify the core component. A precise determination
 of the above offset will then immediately lead to an altitude difference.
 A polarization position angle curve for each component can be constructed
 using the rotating vector model and known $\alpha$ and $\beta$ values.
 A composite of these which reproduces the observed position angle curve
 can be constructed. Then the BCW offsets between intensity and the
 position angle curve centroids for each component will allow
 determination of individual altitudes. It may be noted that the BCW
 offsets can be unambiguously determined only if position angle sweeps
 are clean, which typically occurs for conal components. (If BCW method
 is applied and very large altitudes are considered, then its modification
 to include {\it mfs} will be needed). With the thus obtained conal
 altitude combined with $\Delta_{cc}$ the core altitude can be fixed.
 The filling factors can then be determined by the
 1/3 rule proposed here. Consistency
 of this information with multfrequency observations will
 permit a further look into issues like radius to frequency mapping,
 time delays etc.

 We hope to have demonstrated in previous sections that the kinematical
 effects considered here have the potential to explain the core/cone
 longitude offsets and also shed light on the nature of the emission
 mechanism, and therefore a detailed observational study of these
 offsets holds much promise in elucidating the intricacies
 of pulsar emission.

\end{document}